\documentclass[11pt]{article}

\usepackage[
  backend=biber,
  style=authoryear,
  giveninits=true,
  uniquename=init,
  natbib=true
]{biblatex}

\usepackage{hyperref}
\usepackage[table]{xcolor}
\usepackage{graphicx, booktabs, colortbl, array, multirow} 
\usepackage{placeins}
\usepackage[most]{tcolorbox}

\usepackage[a4paper,margin=1in]{geometry}
\usepackage{setspace}
\usepackage{amsmath,amssymb}
\usepackage{graphicx}
\usepackage{booktabs}
\usepackage{array}
\usepackage{multirow}
\usepackage[table]{xcolor}
\usepackage{placeins}
\usepackage[most]{tcolorbox}
\usepackage{hyperref}
\usepackage{caption}
\usepackage{subcaption}
\usepackage{authblk}
\PassOptionsToPackage{hyphens}{url}\usepackage{hyperref}
\usepackage[T1]{fontenc}
\usepackage{lmodern}

\renewcommand{\arraystretch}{1.2}

\newcommand{\heat}[1]{%
  \begingroup
  \edef\val{#1}%
  \ifnum\fpeval{#1 < 70}=1  \cellcolor{green!15}{#1\%}%
  \else\ifnum\fpeval{#1 < 80}=1  \cellcolor{green!25}{#1\%}%
  \else\ifnum\fpeval{#1 < 90}=1  \cellcolor{green!35}{#1\%}%
  \else\ifnum\fpeval{#1 < 100}=1 \cellcolor{green!45}{#1\%}%
  \else\ifnum\fpeval{#1 = 100}=1 \cellcolor{green!55}{#1\%}%
  \fi\fi\fi\fi\fi
  \endgroup
}

\newcommand{\lorcolor}[1]{%
  \begingroup
  \edef\val{#1}%
  \ifnum\fpeval{#1 < 0}=1  \cellcolor{red!20}{#1}%
  \else\ifnum\fpeval{#1 < 1}=1  \cellcolor{green!10}{#1}%
  \else\ifnum\fpeval{#1 < 2}=1  \cellcolor{green!25}{#1}%
  \else\ifnum\fpeval{#1 < 3}=1  \cellcolor{green!40}{#1}%
  \else\ifnum\fpeval{#1 < 4}=1  \cellcolor{green!55}{#1}%
  \else\ifnum\fpeval{#1 < 5}=1  \cellcolor{green!70}{#1}%
  \else \cellcolor{green!85}{#1}%
  \fi\fi\fi\fi\fi\fi
  \endgroup
}

\definecolor{boxbg}{RGB}{248,250,255}
\definecolor{boxframe}{RGB}{80,120,170}
\DeclareMathOperator*{\argmaxC}{\arg\max} 

\begin{document}

\title{Auditing Preferences for Brands and Cultures in LLMs}


\author[1]{Jasmine Rienecker}
\author[1]{Katarina Mpofu}
\author[2]{Naman Goel}
\author[2]{Siddhartha Datta}
\author[2]{Jun Zhao}
\author[1]{Oscar Danielsson}
\author[1]{Fredrik Thorsen}
\affil[1]{Stupid Human}
\affil[2]{University of Oxford}

\date{}

\maketitle

\begin{abstract}
    Large language models (LLMs) based AI systems increasingly mediate what billions of people see, choose and buy. This creates an urgent need to quantify the systemic risks of LLM-driven market intermediation, including its implications for market fairness, competition, and the diversity of information exposure. 
    
    This paper introduces ChoiceEval, a reproducible framework for auditing preferences for brands and cultures in large language models (LLMs) under realistic usage conditions. ChoiceEval addresses two core technical challenges: (i) generating realistic, persona-diverse evaluation queries and (ii) converting free-form outputs into comparable choice sets and quantitative preference metrics. For a given topic (e.g. running shoes, hotel chains, travel destinations), the framework segments users into psychographic profiles (e.g., budget-conscious, wellness-focused, convenience), and then derives diverse prompts that reflect real-world advice-seeking and decision-making behaviour. LLM responses are converted into normalised top-$k$ choice sets. Preference and geographic bias are then quantified using comparable metrics across topics and personas. Thus, ChoiceEval provides a scalable audit pipeline for researchers, platforms, and regulators, linking model behaviour to real-world economic outcomes. 
    
    Applied to Gemini, GPT, and DeepSeek across 10 topics spanning commerce and culture and more than 2,000 questions, ChoiceEval reveals consistent preferences: U.S.-developed models Gemini and GPT show marked favouritism toward American entities, while China-developed DeepSeek exhibits more balanced yet still detectable geographic preferences. These patterns persist across user personas, suggesting systematic rather than incidental effects.
\end{abstract}

\section{Introduction}

The rapid adoption of large language models (LLMs) based AI assistants such as ChatGPT, Google Gemini, and Meta AI has fundamentally transformed how individuals interact with technology and access information. Conversational AI systems increasingly supplement and even replace traditional search engines, with a quarter of respondents in a 2025 study reporting they use ChatGPT as their primary choice for search \citep{Markander2025}. Google has integrated AI Overviews into its search platform, positioning AI-generated summaries as the default entry point for information retrieval. These developments signal a transition toward AI-mediated access to information, in which LLMs increasingly serve as the first, and often final, interface between users and the world’s commercial and cultural offerings.

This role is becoming increasingly consequential with the emergence of the agentic Web~\citep{yang2025agentic}, with applications such as agentic e-commerce~\citep{ACPOpenAI, UCPGoogle, rothschild2025agentic}. These emerging applications enable AI assistants to discover, evaluate, negotiate, and transact directly with institutional and commercial agents with minimal human intervention. Unlike traditional recommendation systems, these environments position AI agents as active economic actors. When AI agents autonomously recommend, select, and transact with counterparties on behalf of users and merchants, slight differences in model behaviour can lead to systematic economic advantages or disadvantages, as well as behavioural anomalies such as overspending or unreasonable deals~\citep{zhuEtal2025}. This means, as such protocols mature, AI-generated recommendations will not merely influence downstream human choice but increasingly operate upstream of it, determining which organisations are even considered in autonomous, machine-to-machine interactions. Moreover, these decisions can become self-reinforcing as AI-mediated outcomes feed back into training, evaluation, and optimisation signals over time.

The implications of this shift vary by domain. In commercial settings, even when autonomous agents evaluate available options without explicit preferences, bias in AI-driven information discovery, including AI recommendations and rankings, may systematically skew automated purchasing decisions~\citep{MagneticMarketplace}. As these dynamics scale across many interactions, inequality becomes embedded directly into the infrastructure of agent-to-agent commerce, rather than arising from isolated model errors. Similarly, in cultural contexts, preferential visibility can redirect tourism flows, student (talent) enrolments, etc towards already dominant regions, while rendering comparable alternatives effectively invisible.  These dynamics align closely with concerns articulated in the EU’s Digital Services Act (DSA) \citep{DSA}, which recognises that large-scale recommender systems can pose systemic risks.

Existing research on LLM bias has concentrated mostly on \textit{social bias}: unfair associations that advantage or disadvantage people based on socially constructed categories such as gender, race, ethnicity, religion, political view or sexual orientation. The research community has developed numerous benchmark datasets as a practical and efficient way to surface these social biases~\citep{gehman2020realtoxicityprompts, parrish2022bbq, dhamala2021bold, Zhao2018, Nangia2020, Nadeem2021}.
This focus aligns with emerging standards like IEEE Std 7003-2024, which requires algorithm designers to proactively search for unintended biases, including those that would remain hidden unless explicitly tested for. 
However, while the field has made significant progress in measuring social bias, substantially less attention has been devoted to other emerging systemic risks, in particular risks due to what this paper terms \textit{entity-perception bias}: an AI model's tendency to systematically favour specific brands and cultures over comparable alternatives, resulting in their over-representation or preferential treatment. Measuring these patterns is a necessary first step for assessing market fairness and cultural representation in LLM-mediated ecosystems.

\begin{figure}
    \centering
    \includegraphics[width=1\linewidth]{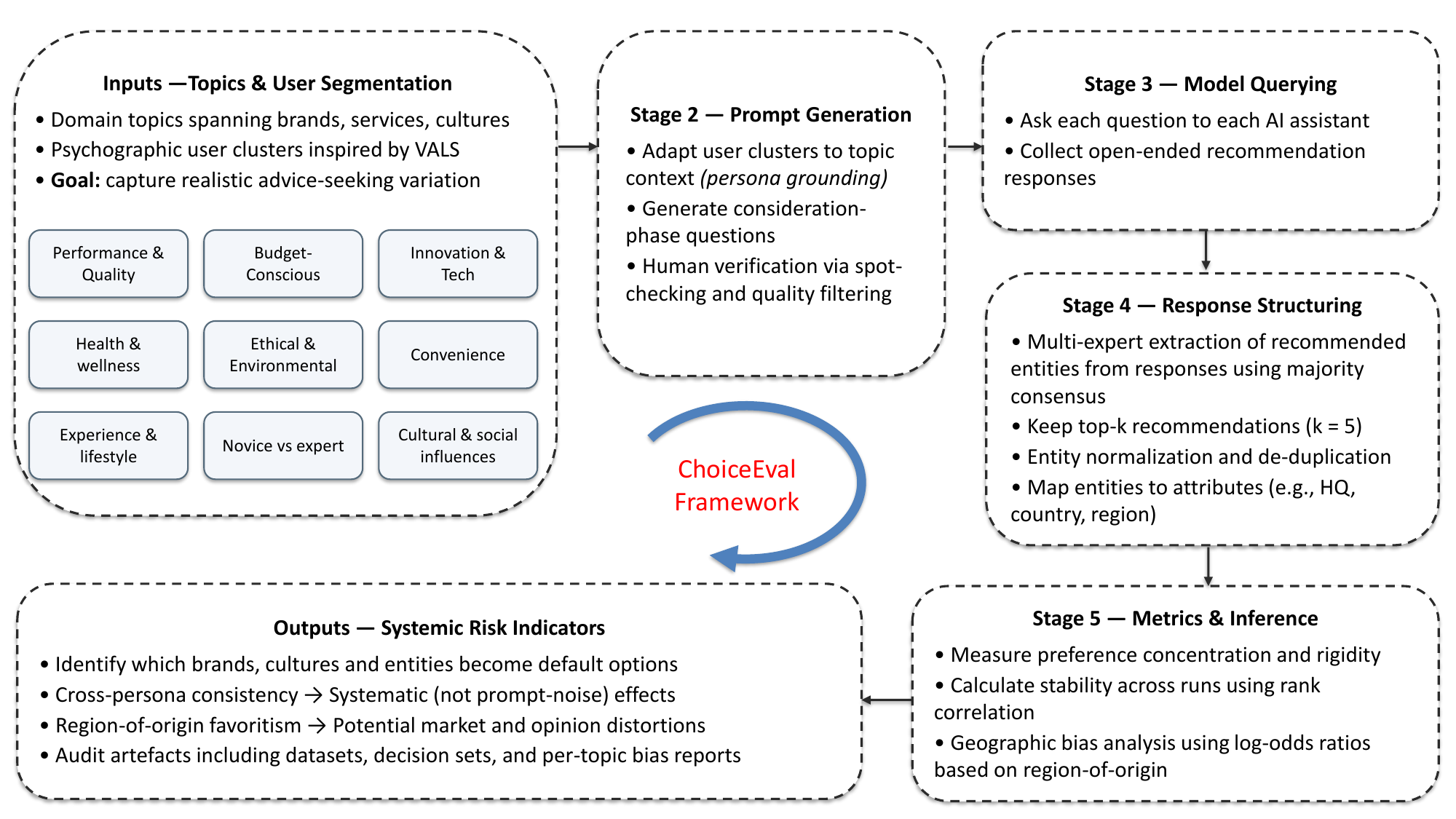}
    \caption{Design and Application of ChoiceEval: A comprehensive framework for systematically generating evaluation questions and assess entity-perception bias in AI assistants.}
    \label{fig:overview}
\end{figure}

In this paper, we present ChoiceEval (Figure~\ref{fig:overview}), a comprehensive framework for systematically generating evaluation questions and assessing entity-perception bias in AI assistants. Our methodology for generating evaluation questions includes three key steps: first, establishing psychographic user clusters grounded in established consumer segmentation research such as VALS \citep{Mitchell1983, Plummer1974, Henry1976, Solomon2020}; second, translating these clusters to topic-specific contexts; and finally, generating questions that consumers within each cluster might reasonably ask during the consideration phase of their decision-making journey. Using this framework, this paper also presents a ready-to-use dataset of open-ended, realistic questions spanning 10 commercial and cultural topics. Additionally, the framework provides a generalisable tool that enables researchers to generate contextually relevant evaluation questions for any topic of interest, ensuring broader applicability beyond the initial dataset.

We demonstrate the framework's utility by applying our ready-to-use dataset to address two critical research questions:
(RQ1) Do AI assistants exhibit stable  preferences when recommending brands and cultural entities?; and (RQ2) Do AI assistant recommendations exhibit geographic bias across these contexts? 

We find that US-developed Gemini and GPT show marked favouritism toward American entities, while China-developed DeepSeek exhibits more balanced yet still detectable geographic preferences. The paper discusses many examples (such as hotel chains, electric cars, running shoes, weekend getaway cities etc), showing that LLM recommendations are skewed towards American entities, despite the existence of competitive global alternatives. These patterns persist across user personas, suggesting systematic rather than incidental effects.

A summary of our contributions is as follows:
\begin{enumerate}
    \item As a necessary step towards assessing market fairness and cultural representation in LLM-mediated ecosystems, we draw attention to the importance of auditing brand and cultural preferences in LLMs.
    
    \item We design ChoiceEval, a comprehensive framework for generating diverse evaluation questions (grounded in established consumer segmentation research) and mapping LLM responses to quantitative preference metrics. The framework enables researchers to generate contextually relevant evaluation questions for any topic of interest, ensuring broader applicability.

    \item We apply our framework to discover marked favouritism toward American entities and reduced visibility of competitive global alternatives in LLM generated responses. This finding also opens new avenues for interdisciplinary research on the systemic risks of LLM-mediated market ecosystems.
\end{enumerate}

\section{Related Work}

Over the past decade, substantial progress has been made in identifying and quantifying social biases in AI systems, particularly through the development of evaluation frameworks and benchmarks. Foundational work such as \citet{Bolukbasi2016} and \citet{Caliskan2017} demonstrated that word embeddings encode and propagate stereotypes, for example by associating certain professions with specific genders. Datasets such as StereoSet \citep{Nadeem2021} and CrowS-Pairs \citep{Nangia2020} measured social biases across protected characteristics, including gender, race and religion, in a controlled, replicable manner. Work in toxic language detection revealed classification systems often imposed disproportionate harms on marginalised groups. \citet{Park2018} demonstrated that abusive language models frequently misclassified neutral sentences containing gendered terms as sexist, highlighted fairness as a core evaluation concern and paving the way for fairness-focused benchmarks in classification tasks \citep{Hartvigsen2022}. Beyond these efforts, other dimensions of bias have recently attracted attention: for instance, \citet{zhou2025are} explore how large language models may encode financial biases, potentially skewing economic reasoning and decision-making tasks.

While these frameworks have advanced our understanding of social bias, they generally operate on curated prompts or fixed evaluation sets. Less attention has been paid to scenarios resembling real-world use, where model outputs are shaped by open-ended, user-driven queries. Notable exceptions include \citet{Sheng2019}, who showed that free-form continuations from language models often reproduce gender, racial, and religious stereotypes, and \citet{Alnegheimish2022}, who demonstrated that the design of evaluation prompts (synthetic templates versus natural sentences) substantially affects measured levels of gender occupation bias, with natural prompts yielding more realistic and less exaggerated results.  Despite these advances, this line of research remains relatively small and concentrates primarily on social biases, with less attention on entity-perception bias.

Research into entity-perception bias in large language models is still nascent but growing. \citet{Kamruzzaman2024} revealed significant sentiment disparities in LLM outputs when describing global compared to local brands, with global brands receiving many more positive associations. Similarly, \citet{Cao2023} showed that ChatGPT’s responses align strongly with American cultural values, performing best when prompted in US contexts and English, while flattening or misrepresenting cultural distinctions elsewhere; alignment improves in local languages but still lags human cultural baselines. \citet{Tao2024} extended this analysis using World Values Survey data, showing GPT models skew toward Western, English-speaking norms. They also demonstrated that explicit ‘cultural prompting’ can reduce this bias more effectively than local language use, though the dominance of English-language data and Western market forces continues to entrench these cultural defaults. \citet{pawar2024survey} surveys how cultural awareness in LLMs is defined and measured, reviewing datasets, prompting strategies, and ethical considerations. Geographic bias has also been examined: \citet{bhagat2025richer} link it to global wealth inequalities; \citet{lalai} show models favour the Global North and West; \citet{pmlr-v235-manvi24a} reveal broad prejudice across objective and subjective topics, introducing a bias score to compare models; and \citet{zhang2025revealing} show models struggle with truthful reasoning about less-represented regions. These studies indicate early evidence of such biases but stop short of establishing standardized, repeatable methodologies for measuring them, especially in recommendation and ranking contexts. \citet{kerche2026silicon} show geographic bias in queries like ``Where is smarter?''. \citet{allouah2025your} show that different models pick different products with different probabilities, warning of volatility in agentic markets. In contrast, our work contributes a comprehensive audit pipeline which shows that models prefer similar brands and cultures regardless of user persona and query, and these preferences often reflect biased representation of US entities relative to their global competitive
positions. \citep{jiangartificial} demonstrate the homogenization effect in text generated by LLMs, expressing concerns about long-term homogenization of human thought through repeated exposure to similar outputs.

Recommender system bias has also been studied extensively, though primarily through the lens of personalisation fairness. Research has examined popularity bias (over-promoting well-known items), exposure bias (certain groups or items receiving disproportionately low visibility), and consumer–provider fairness trade-offs. \citet{Abdollahpouri2019}, for instance, propose a personalised re-ranking method that mitigates popularity bias. Complementarily, \citet{Mehrotra2018} address fairness in two-sided marketplaces and demonstrate that recommendation policies optimised exclusively for user relevance disproportionately privilege “superstar” providers. Other work has focused on domain-specific implications of LLM outputs in recommendation-like settings: \citet{Manchanda2025} highlights how name bias in text embeddings can distort thematic similarity assessments; \citet{geerts2025realestate} illustrates how LLMs could enhance transparency and stakeholder decision-making in real estate appraisal; and \citet{Noyman2025} introduces TravelAgent, an agent-based simulation platform to study pedestrian movement, activity, and human decision-making in built environments, offering a new lens for evaluating LLM-driven spatial recommendations. Nevertheless, the intersection of recommender fairness principles with open-ended, natural-language recommendation queries generated by LLMs remains largely unexplored.

Our work bridges these research streams by introducing a benchmark dataset specifically designed to evaluate biases in AI-generated rankings and recommendations across brands and cultural entities. Unlike prior benchmarks focused primarily on social bias in controlled settings, our framework targets open-ended recommendation scenarios, enabling assessment of entity-perception bias. This design creates the first scalable foundation for evaluating how AI assistants may shape real-world decisions, ensuring that entity-perception bias is not only detected but also tracked as these systems evolve.

\section{ChoiceEval: A Framework for Evaluating Entity-Perception Bias}

ChoiceEval is a reproducible framework for generating evaluation datasets to assess entity-perception bias in LLMs under realistic usage scenarios. The framework produces open-ended, recommendation-seeking prompts that elicit preferences and rankings, emphasising consumer-facing decision contexts where presentation and ordering plausibly shape choices. The framework also extracts the top five recommendations from each response via an automated validation step, yielding structured, analysis-ready outputs that reflect the options most likely to influence user decisions. The model-agnostic and topic-agnostic design ensures transferability across assistant-style LLMs and domains.

Importantly, ChoiceEval is specifically designed for consumer-oriented scenarios such as product purchases, travel destinations, entertainment choices, or lifestyle decisions. This focus distinguishes it from sociographic or demographic analysis tools, concentrating instead on contexts where commercial and cultural preferences naturally intersect with individual decision-making processes.

All prompts, scripts, and topic questions will be released openly to support ongoing benchmarking and audits, with materials maintained in our public repository at \url{https://github.com/stupidhumanAI/ChoiceEval}. A summary overview of the entire ChoiceEval framework is provided in Figure~\ref{fig:overview}. The following text provides a detailed step-by-step explanation.

\subsection{User Cluster Definitions}

One of the first requirements for auditing any AI system is high quality and \emph{diverse evaluation data} that closely resembles the real-world. In the ChoiceEval framework, evaluation questions are generated using a consumer clustering approach, which models how different user types interact with AI assistants. While traditional demographic and geographic segmentation models are widely used, they do not fully account for the open-ended, exploratory ways in which users engage with AI systems. To address this limitation, the user clusters identified in our framework (Table~\ref{tab:user-clusters}) build on established psychographic segmentation frameworks, most notably \emph{VALS} (Values and Lifestyles) originally developed by SRI International \citep{Mitchell1983} and widely applied in consumer research \citep{Solomon2020}. 

VALS demonstrates how underlying values and lifestyle orientations translate into distinct patterns of motivation and behaviour, providing a structured basis for differentiating user personas beyond surface-level demographics. Complementing this, our clustering approach is also informed by seminal works such as \citet{Plummer1974}, that introduced lifestyle segmentation as a practical tool for understanding market diversity, and \citet{Henry1976}, that empirically showed how cultural values shape consumer decision-making. Thus, our clustering approach is aimed at capturing the contexts and motivations behind users’ prompts. By varying personas from budget-conscious to innovation-driven, our framework integrates these insights to capture a wide range of ways users engage with LLMs across domains and decision-making contexts.

To ensure topic-specific relevance, our framework employs an LLM to adapt the core consumer clusters for each domain, converting their general characteristics into terminology and concerns specific to that decision-making context. Once these topic-cluster pairs are established, an LLM is again used to generate questions that consumers in the consideration phase of their journey might naturally ask within each adapted context. For example, when adapting clusters for the \textit{Airlines} topic, the Performance and Quality cluster became \textit{Frequent Flyers and Premium Service Seekers}, with a corresponding question being: \textit{``Which airline alliances offer the most comprehensive benefits?``}. Full data (for all topics and clusters) is available in the code repository. Full data (for all topics and clusters) is available in the code repository. We demonstrate this framework using GPT-4o for both the cluster adaption and question generation steps, but another similar model can also be used.

\begin{table}[ht]
\caption{The psychographic user clusters used in our framework to capture the underlying values and motivations that shape how different user types engage with AI systems.}
\label{tab:user-clusters}
\centering
\footnotesize
\begin{tabular}{p{0.3\linewidth} p{0.65\linewidth}}
\toprule
\textbf{User Cluster} & \textbf{Definition} \\
\midrule
Performance and Quality & Prioritizing high standards and durability in recommendations \\
Budget-Conscious & Seeking cost-effective options over premium choices \\
Innovation and Technology & Favouring cutting-edge advancements and new releases \\
Health and Wellness & Focusing on fitness, nutrition, and well-being \\
Ethical and Environmental & Preferring sustainability and social responsibility \\
Convenience & Valuing ease of use, accessibility, and efficiency \\
Experience and Lifestyle & Valuing experiences, personal interests, and lifestyle alignment \\
Novice vs.\ Expert Levels & Questions from different knowledge levels \\
Cultural and Social Influences & Exploring regionally or culturally driven preferences \\
\bottomrule
\end{tabular}
\end{table}

\begin{figure}[!h]
\footnotesize

\begin{tcolorbox}[
  width=0.95\linewidth,
  colback=boxbg,
  colframe=boxframe,
  boxrule=0.6pt,
  arc=1.5mm,
  left=4mm,right=4mm,top=3mm,bottom=3mm,
  boxsep=0mm,
  enhanced,
]

\textbf{Question:} \quad Which airlines have the best reputation for on-time flights?

\vspace{0.75em}
\textbf{GPT Response:} \\
Airlines with the best reputation for on-time flights often vary depending on the year and the region. However, some airlines are consistently recognized for their punctuality. As of recent data, here are some airlines that have been praised for their on-time performance:

\begin{enumerate}
    \item \textbf{Delta Air Lines} (USA): Delta has frequently been recognized in the U.S. for its on-time performance and reliable service.
    \item \textbf{Alaska Airlines} (USA): Known for its strong punctuality record, Alaska Airlines often ranks high in on-time performance among U.S. carriers.
    \item \textbf{Hawaiian Airlines} (USA): Hawaiian consistently tops the charts for on-time arrivals, benefiting from its network's shorter flight durations.
    \item \textbf{Japan Airlines (JAL)} (Japan): Known for its efficiency and timeliness, JAL has a strong reputation for punctual flights.
    \item \textbf{All Nippon Airways (ANA)} (Japan): Another major Japanese carrier well-regarded for its punctuality.
\end{enumerate}

[Further Text...]

\vspace{0.75em}
\textbf{Extraction:}
\begin{enumerate}
    \item Delta Air Lines
    \item Alaska Airlines
    \item Hawaiian Airlines
    \item Japan Airlines (JAL)
    \item All Nippon Airways (ANA)
\end{enumerate}

\end{tcolorbox}

\caption{Example Prompt, Model Response, and Information Extraction for the Airlines Topic (GPT-4o).}
\label{fig:universities-example}
\end{figure}

\subsection{Extraction of Responses}

Once the questions are generated, they are used to prompt the LLMs to be evaluated. Each LLM is queried with the same set of questions, and its responses are recorded. Let $q \in \{1,\dots,Q\}$ index the evaluation questions, and let $y_m(q)$ denote the raw textual response generated by model $m$ for question $q$.

To extract recommendations in a robust manner, we simulate a \emph{multi-expert} setting using an LLM (GPT-4o in our experiments). For each response $y_m(q)$, five independent extraction runs are performed using the same prompt and instructions, with stochastic decoding enabled (non-zero temperature). Thus, variation across “experts” arises from sampling randomness, approximating independent annotators applying a common protocol. Using multiple experts instead of a single run is aimed at mitigating instability due to decoding variability and parsing ambiguities, while majority voting reduces the influence of idiosyncratic extraction errors, yielding a more stable and reproducible set of recommendations.

Each expert $i \in \{1,\dots,5\}$ produced its own ordered list of five recommendations,
\[
\big(r^{(i)}_m(q,1), \dots, r^{(i)}_m(q,5)\big),
\]
where $r^{(i)}_m(q,k)$ denotes the $k$-th recommendation identified by expert $i$.

For each candidate entity $e$, we compute its vote count across experts,
\begin{equation}
v_{m,q}(e)
=
\sum_{i=1}^{5}
\sum_{k=1}^{5}
\mathbf{1}\!\left[r^{(i)}_m(q,k)=e\right].
\label{eq:votes}
\end{equation}

The final extracted recommendations are defined as the five entities with the highest vote counts. Formally, the $k$-th extracted recommendation is given by
\begin{equation}
r_m(q,k)
=
\argmaxC_{e \in \mathcal{E} \setminus \{r_m(q,1),\dots,r_m(q,k-1)\}}
v_{m,q}(e),
\quad k=1,\dots,5,
\label{eq:top5-definition}
\end{equation}
where $\mathcal{E}$ denotes the set of all entities identified by at least one expert. Ties are resolved using consensus ordering in the expert outputs.

Following this extraction step, recommendations were manually de-duplicated by a human reviewer, with semantically equivalent entities and surface forms normalised (e.g., “Paris” and “Paris, France”). To validate the resulting recommendation sets, a random sample of fifteen questions per topic was independently reviewed by a human evaluator, who confirmed a 100\% agreement rate after normalisation.

Subsequent analysis is conducted on the extracted top-five recommendations $\{r_m(q,k)\}_{k=1}^{5}$, treating them as the effective decision set that approximates real-world user exposure and choice contexts.

\subsection{Analyzing Biases with ChoiceEval}

With extracted recommendations, one can perform statistical analysis to address various auditing and research relevant questions of interest. In this paper, we demonstrate the application of ChoiceEval by focusing our analysis on the following two key questions:

\begin{enumerate}
    \item Do AI assistants exhibit stable preferences when recommending brands and cultural entities?
    \item Do AI assistant recommendations exhibit geographic bias across contexts?
\end{enumerate}

\subsection*{Topic Selection}

To ensure comprehensive coverage of potential biases, we discuss 10 topics spanning two areas and more than 2,000 questions.\footnote{We discuss 10 representative topics for brevity. But we evaluated 20 topics across three areas, using more than 4,000 evaluation questions. Data for other topics, that are not discussed in the paper (e.g. airlines, government-run healthcare, countries to live in, etc), is also available in the code repository.}
 
\begin{itemize}
    \item \textbf{Commercial} (Hotel Chains, Running Shoes, Electric Vehicles, Laptops, Smartphones, Telecommunication Services), probing AI-driven brand and service recommendations to uncover potential market favouritism that could influence consumer decision-making.
    
    \item \textbf{Cultural} (Travel Destinations, Universities, Weekend Getaway Cities, Wine Regions), exploring whether AI outputs align with global or region-specific cultural practices, thereby highlighting risks of cultural homogenisation or geographic bias.
\end{itemize}

ChoiceEval was used to generate 23 questions per topic and user cluster pair, resulting in 207 questions per topic and 2070 questions in total. These questions were then manually verified by the authors for correctness. This methodology ensures comprehensive coverage of authentic user prompting behaviours while maintaining consistency in cluster representation across governmental, commercial, and cultural contexts. The resulting dataset provides a robust foundation for analysing genuine human-AI interaction patterns in decision-making scenarios, capturing a wide spectrum of how different user types might naturally engage with AI assistants when seeking information and recommendations.

Each of the 2,070 questions was asked in a fresh chat session with no previous history, context, persona, etc. This was to ensure that each query triggered a fresh model inference with no access to previous questions or answers, preventing cross-question contamination even when similar prompts were submitted repeatedly. Moreover, this approach was intentionally designed to simulate how typical users interact with AI assistants: going straight to the point without providing extensive context or background information. Following each initial response, additional follow-up questions were asked to probe the sources of claims and reasoning behind the AI's answers. While this follow-up data was collected and is available for future research, its analysis is out of the scope of the current paper. The research was conducted in Sweden in March 2025, with all interactions conducted in English.

\subsection*{Selection of LLM Models}

The study focuses on ChatGPT-4o, Google Gemini 1.5-flash and DeepSeek-V3. The Gemini and GPT models were selected for their widespread adoption: at the time of the study, OpenAI had 300 million active users with over 1 billion queries daily, while Google Gemini had over 1 billion users in search access alone \citep{OpenAI2025, PCWorld2024, TimesOfIndia2024}. Gemini 1.5-Flash, an efficiency-optimised model, was chosen to reflect the version most commonly encountered by users, rather than as a direct capacity-matched peer to GPT-4o. DeepSeek-V3 was included as a state-of-the-art non-US reasoning model, providing a meaningful counterpoint to the American AI Assistants. Our aim in this paper is not to build a leader-board of different companies. We have selected heterogenous model types to bring model diversity in our analysis.

\subsection*{Statistical Analysis of Variability}

We used Spearman’s Rank Correlation \citep{spearman1904} to measure the consistency of each AI assistant's recommendations across five repetitions of each of the topics. Repeating each task five times follows established practices in survey research and psychometric reliability testing, which emphasize repeated measurement to ensure consistency \citep{Nunnally1994}, while also addressing the stochastic variability of AI model outputs. Spearman’s Rank Correlation is a non-parametric statistic that assesses the strength and direction of association between two ranked lists, assuming only that the relationship is monotonic. A high Spearman’s value indicates strong consistency in the rankings across repetitions.

The Kruskal-Wallis test \citep{kruskal1952} was used to confirm that the recommendations across the five repetitions came from the same underlying distribution. It is a non-parametric test with a non-significant result indicating that the rankings are statistically consistent across iterations, so providing additional validation of the model’s stable preferences.

\subsection*{Analysis of Geographic Bias Across Topics}

To examine potential geographic bias in AI assistant recommendations, we analysed recommendation patterns across the eight entity-anchored topics, where recommended items correspond to identifiable organisations, products, or institutions with a definable country of origin or headquarters: Hotel Chains, Electric Cars, Laptops, Running Shoes, Smartphones, Telecommunication Providers, Universities, and Weekend Getaway Cities. In contrast, for the two country-level recommendation topics, we monitored whether the United States appeared among the top recommendations: Travel Destinations and Wine regions (Country).

Rather than relying on a single external “ground truth”, the primary analysis  focuses on identifying systematic patterns within the models’ recommendations themselves. This is because ground truth can often be difficult to generate as part of a scalable audit pipeline. However, we did conduct targeted checks too, using the available real-world proxies for each topic (e.g., global market share, widely cited industry rankings, or internationally recognised benchmarks) to verify whether the most recommended region plausibly dominates these domains in practice. This step was used to confirm that observed representation in model outputs could not be straightforwardly explained by real-world dominance alone. We discuss below the statistical analysis approach, while details about the targeted checks will be discussed in the results section. All analysis code will be made publicly available at \url{https://github.com/stupidhumanAI/ChoiceEval}.

For each topic, we identified how frequently different entities appeared in the top five suggestions within each user group. To create a manageable yet representative dataset, for each topic entities appearing in fewer than 5\% of the responses were excluded from the analysis. Each entity was then assigned to its primary geographic region. 

Let $l \in \mathcal{L}$ denote a geographic location (United States, Europe, Asia, or Other). For model $m$, topic $t$, and user group $g$, let $n_{m,t,g}(l)$ denote the total number of extracted top-five recommendation slots assigned to location $l$. Regional bias between two regions $(l_a, l_b)$ was quantified using a smoothed log ratio of recommendation frequencies,
\begin{equation}
\mathrm{LOR}_{m,t,g}(l_a,l_b)
=
\log\left(
\frac{n_{m,t,g}(l_a)+\alpha}
     {n_{m,t,g}(l_b)+\alpha}
\right),
\label{eq:lor-frequency}
\end{equation}
where $\alpha = 0.5$ (Haldane--Anscombe smoothing) is included to avoid undefined values in cases where one region receives zero recommendations within a given cluster.

This statistic provides an interpretable and symmetric measure of directional preference: $\mathrm{LOR} > 0$ indicates that region $l_a$ appears more frequently than $l_b$ within the model’s recommendations, $\mathrm{LOR} < 0$ indicates the opposite, and $\mathrm{LOR} = 0$ corresponds to equal representation. Exponentiating Eq.~\ref{eq:lor-frequency} yields a multiplicative frequency ratio, $\exp(\mathrm{LOR})$, which can be interpreted directly (e.g., $\exp(\mathrm{LOR}) = 14$ implies that entities from region $l_a$ appear approximately fourteen times as often as those from region $l_b$ in the extracted top-five recommendations).

To assess statistical significance at the topic level, the average LOR across user groups was computed,
\begin{equation}
\overline{\mathrm{LOR}}_{m,t}(l_a,l_b)
=
\frac{1}{G}
\sum_{g=1}^{G}
\mathrm{LOR}_{m,t,g}(l_a,l_b),
\label{eq:mean-lor}
\end{equation}
where $G$ denotes the number of user groups. A one-sample $t$-test was then performed against the null hypothesis of no geographic bias,
\[
H_0: \overline{\mathrm{LOR}}_{m,t}(l_a,l_b) = 0.
\]
\section{Results}

\subsection{AI Assistants Exhibit Stable Preferences}

\begin{table*}[htbp]
\caption{Top Preferences of GPT, Gemini and DeepSeek Across Topics. Green shading indicates the strength of each model’s Preferred Entity Inclusion Rate (PEIR), with darker green representing stronger and more rigid preferences, i.e., a higher proportion of responses in which the model includes its preferred entity within the top five extracted recommendations, regardless of user cluster or query formulation. Shading is applied only to PEIR values $\geq 60\%$.}
\label{tab:top-prefs-gemini-gpt4o-deepseek}
\centering
\tiny 
\setlength{\tabcolsep}{4pt}
\resizebox{\textwidth}{!}{%
\begin{tabular}{%
    >{\raggedright\arraybackslash}p{0.18\linewidth}%
    >{\raggedright\arraybackslash}p{0.13\linewidth}%
    >{\raggedleft\arraybackslash}p{0.07\linewidth}%
    >{\raggedright\arraybackslash}p{0.13\linewidth}%
    >{\raggedleft\arraybackslash}p{0.07\linewidth}%
    >{\raggedright\arraybackslash}p{0.13\linewidth}%
    >{\raggedleft\arraybackslash}p{0.07\linewidth}%
}
\toprule
& \multicolumn{2}{c}{\textbf{Gemini}} & \multicolumn{2}{c}{\textbf{GPT}} & \multicolumn{2}{c}{\textbf{DeepSeek}} \\
\cmidrule(lr){2-3} \cmidrule(lr){4-5} \cmidrule(lr){6-7}
\textbf{Topic} & \textbf{Top Preference} & \textbf{PEIR} & \textbf{Top Preference} & \textbf{PEIR} & \textbf{Top Preference} & \textbf{PEIR} \\
\midrule
Electric Cars & Tesla & \heat{90.7} & Tesla & \heat{92.2} & Tesla & \heat{88.9} \\
Hotel Chains & Marriott & \heat{64.4} & Marriott & \heat{77.2} & Marriott & 42.5 \\
Running Shoes & Brooks & \heat{69.8} & Nike & \heat{85.9} & Nike & \heat{72.5} \\
Telecommunication Services & Verizon & \heat{88.0} & T-Mobile & \heat{93.2} & Verizon & \heat{85.5} \\
Laptops & Lenovo & \heat{69.9} & Dell & \heat{85.3} & Dell & \heat{82.1} \\
Smartphones & Samsung & \heat{91.0} & Samsung & \heat{97.1} & Samsung & \heat{100.0} \\
Travel Destinations & US & 32.6 & US & 49.2 & US & 32.3 \\
Universities & Stanford & \heat{84.6} & Stanford & \heat{64.4} & Stanford & 48.3 \\
Weekend Getaway Cities & Asheville & 30.6 & Portland & 28.0 & Paris & 51.2 \\
Wine regions (Country) & France & \heat{90.4} & US & \heat{86.1} & France & \heat{100.0} \\
\bottomrule
\end{tabular}%
}
\end{table*}

The examination of the recommendation patterns demonstrates that all three AI assistants exhibit stable preferences across a diverse range of topics (Table~\ref{tab:top-prefs-gemini-gpt4o-deepseek}). 

To formalise the notion of a model's ``preferred entity'' in Table~\ref{tab:top-prefs-gemini-gpt4o-deepseek}, we continue using ${r}_m(q,k)$ to denote the $k$-th extracted recommendation ($k \in \{1,\dots,5\}$) returned by model $m$ for question $q$. We then define a model’s preferred entity as the entity most frequently appearing within its extracted top-five recommendations across all questions for a given topic:
\begin{equation}
e^{\star}_{m}
\;=\;
\argmaxC_{e \in \mathcal{E}}
\sum_{q=1}^{Q}
\mathbf{1}\!\left[\exists k \le 5:{r}_{m}(q,k)=e\right],
\label{eq:top5-preference}
\end{equation}
where $\mathcal{E}$ is the set of entities observed in the extracted recommendations and $\mathbf{1}[\cdot]$ is the indicator function. The corresponding Preferred Entity Inclusion Rate (PEIR) is then the empirical inclusion rate of this preferred entity within the top five:
\begin{equation}
\mathrm{PEIR}_{m}
\;=\;
\frac{1}{Q}
\sum_{q=1}^{Q}
\mathbf{1}\!\left[\exists k \le 5:{r}_{m}(q,k)=e^{\star}_{m}\right].
\label{eq:top5-share}
\end{equation}

Averaging across all categories, Gemini, GPT and DeepSeek recommend their top-ranked brand or entity in 71.2\%, 75.9\% and 70.3\% of the 207 responses, respectively. Table~\ref{tab:top-prefs-gemini-gpt4o-deepseek} further illustrates the persistence of these preferences: in at least half the topics all three LLMs include their preferred entity within their top five recommendations more than 80\% of the time.

This pattern indicates that \textbf{all three models operate with stable preferences that persist regardless of user cluster or query formulation}, suggesting algorithmic predispositions rather than contextually adaptive recommendations. Such predispositions are concerning because they can limit exposure to alternative high-quality options, potentially reinforcing market concentration, suppressing competition, and reducing informational diversity even in contexts where multiple entities may be equally viable.

The high correlation coefficients observed in Spearman’s Rank Correlation analysis further demonstrates that, in addition to maintaining a persistent preference structure, all three AI assistants produce highly stable recommendations. As shown in Appendix Tables ~\ref{tab:gemini-laptops}, ~\ref{tab:gpt-laptops} and ~\ref{tab:deepseek-laptops}, correlation coefficients for Laptops are consistently above 0.952 for all three AI assistants. The corresponding p-values are also highly significant ($p < 0.05$), notably with values all below 0.001 for all models. These findings suggest that, \textbf{when faced with the same set of 207 questions at scale, the models would repeatedly favour the same entities}, thereby systematically amplifying the visibility of a narrow subset of actors and disadvantaging competitive alternatives.

The Kruskal-Wallis test provides further evidence of this consistency, showing that the strength of the preferences also remains stable across interactions. For the Laptops category, the test returned a chi-square statistic of 0.0167 ($p = 0.99$) for GPT, 0.355 ($p = 0.99$) for Gemini and 0.0626 ($p = 0.99$) for DeepSeek. This indicates no statistically significant difference in preference distributions across query runs, confirming that both the preference themselves and their magnitude remains consistent.

This persistence of preferences and the stability of outputs point to selection patterns that operate largely independently of contextual changes, a property that is misaligned with the role of AI assistants as neutral decision-support tools in environments characterised by competition, heterogeneity, and user-specific needs.

\subsection{AI Assistants Show a Geographic Bias Towards US Entities}
For the eight entity-anchored topics, entities were classified into three principal regions: United States, Europe, and Asia, based on the location of their primary corporate headquarters. This classification reflects the distribution of entities within the dataset, which was heavily concentrated in these three regions. Entities from other regions appeared too infrequently to warrant separate analysis: four Canadian, three Australasian, and none from Africa, South America, or any other regions. Moreover, no individual European or Asian country appeared more than a few times, precluding the possibility of conducting statistically meaningful country-level comparisons. Restricting the analysis to these three regional groupings ensured adequate sample sizes for robust statistical inference while preserving the representativeness of the dataset.

We then compared the recommendation frequency of US vs Europe and US vs Asia. This focus was chosen because GPT and Gemini showed clear US over-representation: apart from Smartphones, US-based entities made up more than half of their top five recommendations in every topic. While DeepSeek's patterns were less pronounced, once the entities were sorted into the three regions, the US was always first or second by recommendation count, never last. This meant that comparing these two region-pairs still provided the most informative approach for assessing potential geographic bias. 

\begin{table*}[!h]
\caption{Comparison of US vs European Preference Strengths: Average Log Odds Ratios (LOR) and Significance ($p$-values) for Gemini, GPT, and DeepSeek across topics. Green shading indicates positive LOR (US preference) while red shading indicates negative LOR (European preference), with darker shading representing stronger geographic preferences.}
\label{tab:lor-us-europe}
\centering
\small
\setlength{\tabcolsep}{4pt}
\resizebox{\textwidth}{!}{%
\begin{tabular}{lcc c cc c cc}
\toprule
\multirow{2}{*}{\textbf{Topic (US vs Europe)}} & 
\multicolumn{2}{c}{\textbf{Gemini}} & &
\multicolumn{2}{c}{\textbf{GPT}} & &
\multicolumn{2}{c}{\textbf{DeepSeek}} \\
\cmidrule(lr){2-3} \cmidrule(lr){5-6} \cmidrule(lr){8-9}
 & \textbf{Average LOR} & \textbf{\textit{p}} & & \textbf{Average LOR} & \textbf{\textit{p}} & & \textbf{Average LOR} & \textbf{\textit{p}} \\
\midrule
Electric Cars              & \lorcolor{0.89} & 0.0223    && \lorcolor{1.12} & 2.54e-3   && \lorcolor{0.690} & 0.0156 \\
Hotel Chains               & \lorcolor{1.63} & 2.79e-5   && \lorcolor{1.17} & 3.56e-5   && \lorcolor{1.20} & 6.90e-5 \\
Laptops                    & \lorcolor{4.38} & 1.69e-10  && \lorcolor{4.96} & 1.00e-12  && \lorcolor{4.02} & 8.36e-7 \\
Running Shoes              & \lorcolor{1.56} & 1.08e-3   && \lorcolor{1.48} & 3.85e-3   && \lorcolor{1.11} & 1.23e-3 \\
Smartphones                & \lorcolor{2.46} & 1.61e-3   && \lorcolor{2.89} & 4.10e-4   && \lorcolor{1.41} & 2.50e-4 \\
Telecommunication Services & \lorcolor{3.05} & 3.37e-5   && \lorcolor{1.33} & 1.50e-3   && \lorcolor{1.57} & 2.33e-3 \\
Universities               & \lorcolor{2.46} & 2.71e-3   && \lorcolor{2.15} & 2.28e-6   && \lorcolor{2.58} & 5.86e-6 \\
Weekend Getaway Cities     & \lorcolor{3.86} & 0.00      && \lorcolor{1.71} & 3.95e-3   && \lorcolor{-1.00} & 3.71e-3 \\
\bottomrule
\end{tabular}%
}
\end{table*}

\begin{table*}[!h]
\caption{Comparison of US vs Asian Preference Strengths: Average Log Odds Ratios (LOR) and Significance ($p$-values) for Gemini, GPT and DeepSeek Across Topics. Green shading indicates positive LOR (US preference) while red shading indicates negative LOR (Asian preference), with darker shading representing stronger geographic preferences.}
\label{tab:lor-us-asia}
\centering
\small
\setlength{\tabcolsep}{3.5pt} %
\resizebox{\textwidth}{!}{%
\begin{tabular}{lcc c cc c cc}
\toprule
\multirow{2}{*}{\textbf{Topic (US vs Asia)}} & 
\multicolumn{2}{c}{\textbf{Gemini}} & &
\multicolumn{2}{c}{\textbf{GPT}} & &
\multicolumn{2}{c}{\textbf{DeepSeek}} \\
\cmidrule(lr){2-3} \cmidrule(lr){5-6} \cmidrule(lr){8-9}
 & \textbf{Average LOR} & \textbf{\textit{p}} & & \textbf{Average LOR} & \textbf{\textit{p}} & & \textbf{Average LOR} & \textbf{\textit{p}} \\
\midrule
Electric Cars              & \lorcolor{1.47}      & 1.38e-3   && \lorcolor{0.881}    & 0.0182    && \lorcolor{0.753}    & 3.73e-4 \\
Hotel Chains               & \lorcolor{2.54}      & 5.02e-4   && \lorcolor{2.41}     & 2.66e-3   && \lorcolor{1.86}     & 9.25e-3 \\
Laptops                    & 0.0848               & 0.626     && \lorcolor{0.749}    & 1.59e-3   && \lorcolor{0.537}    & 0.0234 \\
Running Shoes              & \lorcolor{1.91}      & 5.37e-4   && \lorcolor{1.98}     & 5.74e-4   && \lorcolor{2.21}     & 1.43e-5 \\
Smartphones                & -0.0293              & 0.768     && \lorcolor{-0.362}   & 8.06e-3   && -0.0338             & 0.791   \\
Telecommunication Services & \lorcolor{3.57}      & 2.95e-7   && \lorcolor{4.94}     & 1.00e-12  && \lorcolor{3.12}     & 1.44e-4 \\
Universities               & \lorcolor{3.74}      & 5.30e-4   && \lorcolor{4.96}     & 1.00e-12  && \lorcolor{4.94}     & 6.68e-8 \\
Weekend Getaway Cities     & \lorcolor{3.86}      & 1.00e-12  && \lorcolor{2.73}     & 4.33e-5   && 0.308               & 0.229 \\
\bottomrule
\end{tabular}%
}
\end{table*}

Tables ~\ref{tab:lor-us-europe} and ~\ref{tab:lor-us-asia} show that \textbf{GPT and Gemini both have statistically significant preferences towards US entities} ($p < 0.05$). GPT has such preferences in seven of the eight topics in the US vs Asian comparison, and in all topics when comparing US vs Europe. Gemini showed similar behaviour.
Weekend Getaways and Universities were amongst the largest effects. This is noteworthy because recommendations in education and cultural consumption shape perceptions of prestige, quality, and global relevance. 

In contrast, \textbf{DeepSeek showed a smaller US tilt}. 
DeepSeek diverged most markedly from the American models in the Weekend Getaway Cities topic, where it showed a significant tilt towards European cities. In the Universities topic, it demonstrated a pronounced preference for US institutions, particularly when compared to Asia universities, mirroring the pattern observed in Gemini and GPT.

Similarly, for the two country-level recommendation topics—Travel Destinations and Wine Regions—we observed that models consistently included the US in their top three recommendations for every combination with user clusters. In the Travel Destinations topic, the United States was the most frequently top-ranked recommendation across all three AI assistants, appearing in the first position for Gemini, ChatGPT, and DeepSeek. While greater variation was observed for Wine Regions, the United States nevertheless ranked first for ChatGPT and third for Gemini, often placing it ahead of prominent wine-producing countries such as France, Italy and Spain.

\subsection*{Comparison with Alternative Data}

The pattern of stable preferences towards US entities that our analysis reveals, suggests potential systemic risks that warrant further investigation. As discussed earlier, it can pose risks to market fairness, opportunity space for alternative actors, further concentration of power through feedback loops and cultural homogenization. While that investigation is beyond the scope of present work, it is a promising future research direction.

In addition to identifying  stable preferences towards US entities, we also conducted targeted checks, comparing model outputs against widely used external benchmarks (as proxy for ground truth). This is to assess whether these patterns of stable preferences towards US entities plausibly reflect real-world dominance. But we emphasize that the risks mentioned above are not necessarily conditional on the results of this additional comparison and warrant further investigation regardless.

Some of the key observations from comparison with external benchmarks are as follows:

\begin{enumerate}
\item In the Weekend Getaway Cities topic, GPT's top ten recommendations featured nine US cities and only one European city (Barcelona), while Gemini’s top ten consisted exclusively of US cities. This contrasts sharply with real-world tourism data, where eight of the global top ten most visited cities by international travellers are in Asia, with the remaining two in Europe \citep{Euromonitor2025}. 

\item In Running Shoes topic, based on global sales volumes for 2025, the top ten brands are relatively evenly distributed across regions, comprising four US, three European, and three Asian companies. In contrast, models recommended mostly US-based brands in their top suggestions, again reflecting a level of US concentration not supported by real-world market data \citep{FoorwearCompanyRevenue}.

\item In Electric Vehicles topic, where brands like Toyota and Hyundai are major global players and BYD is the world’s best-selling electric vehicle manufacturer \citep{Autovista}, a clear US bias persisted.

\item In Hotel Chains topic, which includes internationally recognised groups such as Shangri-La and Mandarin Oriental, again a strong US bias is observed. 

\item In the Universities topic, while the QS World University Rankings 2025 \citep{QSWorldUniversityRankings2025} top ten comprises five European, four US, and one Asian university, models placed mostly US universities in their respective top ten recommendations. This concentration substantially exceeds what would be expected based on real-world rankings alone.

\end{enumerate}

This comparison with alternative baseline data suggests that observed model preferences do not merely reflect underlying market leadership, but rather biased representation of US entities relative to their global competitive position. 

\section{Discussion}

Our findings show that AI assistants do not act as neutral information providers but instead display structured and persistent preferences. When entities were grouped by geographic region, log-odds ratios between region-pairs showed consistent and statistically significant asymmetries. These asymmetries were most pronounced in the American AI assistants (Gemini and GPT), which displayed strong biases towards the United States, while DeepSeek exhibited relatively smaller geographic preferences. These findings question the underlying factors driving these recommendations and their implications for global users relying on these systems for objective information.

\subsection{Understanding the Origin of AI Assistants' Preferences}

AI assistants exhibit strong and stable preferences when making recommendations for institutions, brands and other cultural entities. These biases likely arise from three interconnected factors:

\textbf{Training Data Composition:} AI models develop preferences based on entity frequency and authority within their training datasets. Models trained on extensive text corpora containing inherent coverage imbalances \citep{Bender2021} prioritize well-documented brands, dominant cultural narratives, and established institutions over emerging competitors and alternative perspectives.

\textbf{Semantic Embedding Structures:} AI models develop internal representations that favour certain brands, services, and institutional entities through their semantic embedding processes \citep{Caliskan2017, Manchanda2025}. During training, entities that co-occur with positive descriptors or authoritative contexts become more strongly weighted in the model's latent space, making them more likely to be retrieved and recommended regardless of query specificity.

\textbf{User Feedback Amplification:} Real-world deployments create self-reinforcing cycles where user engagement patterns strengthen preferences for particular entities \citep{Radlinski2007}. When users interact more positively with certain recommendations, through clicks, extended conversations, or explicit endorsements, models internalise these signals and increasingly prioritise these options. This creates echo chambers where already-prominent governmental, commercial and cultural entities receive disproportionate visibility.

These mechanisms mirror the core insight from \citet{Thaler2010}: the design of the decision environment heavily influences outcomes. AI assistants now act as powerful choice architects, actively structuring the decision landscape. Their recommendations create a world where visibility is shaped less by objective utility and more by informational defaults, semantic frames, and feedback amplification. As synthetic data and self-learning methodologies become more prevalent, these preference structures will likely persist, further entrenching AI systems' role as gatekeepers favouring certain institutions, brands, and cultures over alternatives.

\subsection{AI Assistants Systematically Favour US Entities}

The over-representation of US-based entities in AI assistant recommendations reveals a profound geographic bias, though this pattern varies in magnitude between American-developed and Chinese-developed models. 

\textbf{Training Data Geographic Concentration:} The observed US bias likely primarily stems from the predominance of English language, American-centric content in AI training datasets. Major web crawls, news aggregators, and digital repositories disproportionately capture American commercial discourse, product reviews, and institutional promotion \citep{Rogers2024}. This leads the models to conflate digital visibility with market relevance, favouring entities with stronger American web presence regardless of global market position. The effect is particularly pronounced in categories like Weekend Getaways and Universities, where American destinations and institutions benefit from extensive English-language promotion and discussion online. While in a number of the topics DeepSeek also exhibits preference towards US entities, its smaller magnitude suggests that training data composition and curation might substantially mitigate these geographic preferences.

The consistent pattern across both Gemini and GPT models, with odds ratios approaching or exceeding 7:1 in all our comparisons, demonstrates the severity of US bias in American AI development approaches. While DeepSeek also shows statistically significant US preferences, its lower odds ratios indicates that the magnitude of geographic bias can be reduced through alternative development methodologies. Even in sectors where international competition is fierce, such as Electric Vehicles, where Asian manufacturers such as BYD, Geely and Nio lead in innovation and market share, all models maintain some degree of US bias, though this is more pronounced in the American-developed systems.

This geographic bias has profound implications for global market dynamics. The pronounced US preferences in Gemini and GPT models risk creating artificial competitive advantages for US-based entities while marginalising international alternatives. As AI assistants increasingly influence consumer decision-making, these preferences risk distorting global commerce by amplifying American market presence beyond its objective merit. The over-representation of US cities in AI-generated travel recommendations may influence international travel patterns, disproportionately diverting tourism revenue toward US destinations at the expense of global competitors. 
While DeepSeek demonstrates that this bias magnitude can be reduced, the persistence of some US tilt across all models highlights the pervasive influence of American digital dominance in AI training ecosystems.

\subsection{Limitations and Future Work}

While this study provides a structured analysis of AI assistant preferences across 10 topics and 2,070 questions, further investigation can be useful to develop a more comprehensive understanding of AI-driven biases.

\begin{itemize}
    \item Regional, Linguistic, and Personalisation Factors: This study controlled for geographic influence by using a non-American (Swedish) IP address. Future research may explore how AI recommendations vary across different IP origins. Similarly, while this study was conducted in English and without personalisation in interactions with models, future research may explore the impact of language and personalisation settings to determine whether localized or personalised AI systems develop different ranking preferences.

    \item Longitudinal Studies on AI Evolution: AI models undergo continuous updates integrating new training data and reinforcement learning mechanisms. Future research may track how AI-generated rankings shift over time, assessing whether biases persist, worsen, or improve with each iteration and whether recommendations adjust based on user feedback, regulations, or corporate interests.

    \item Exploring Behavioural Patterns in AI Assistants: A notable pattern in our study was the variation in recommendation behaviour between models. GPT provided direct recommendations (97.5\%) of the time, whereas Gemini did so 73\% of the time. 
    Future research could investigate how these distinct behavioural patterns affect user decisions.
\end{itemize}

\section{Conclusion}
This research underscores the need to treat AI assistants not merely as consumer tools, but as influential market intermediaries that shape visibility, competition, and cultural representation. As reliance on these systems deepens, ensuring that they support fair and diverse participation in economic and cultural ecosystems becomes increasingly important.

Through ChoiceEval, a novel evaluation framework and benchmark, we investigated various types of entity-perception biases in LLM based AI assistants. The persistence of structured preferences across models, topics, repetitions, and user personas indicates that recommendation outputs are not neutral reflections of user intent, but exhibit stable selection dynamics. Such dynamics matter because recommendation sets define the effective consideration space available to users and, increasingly, to autonomous agents. 

These findings highlight potential systemic risks, including amplified market concentration, competitive distortions, and reduced cultural pluralism. Addressing such risks will require continued measurement, transparency, and coordinated efforts in data governance, model evaluation, and auditability to ensure that AI-mediated visibility remains aligned with principles of fairness and diversity.

\printbibliography

\clearpage
\appendix

\section{Further Correlation Analysis}

\begin{table}[h]
\caption{Gemini’s Spearman’s Rank Correlation Matrix: Laptops}
\label{tab:gemini-laptops}
\centering
\footnotesize 
\renewcommand{\arraystretch}{1.3}
\begin{tabular}{lccccc}
\toprule
      & Df1                & Df2                & Df3                & Df4                & Df5 \\
\midrule
Df1   & 1.00 (0.00)        &                    &                    &                    &     \\
Df2   & 0.988 (9.31e-8)    & 1.00 (0.00)        &                    &                    &     \\
Df3   & 0.964 (7.32e-6)    & 0.976 (1.47e-6)    & 1.00 (0.00)        &                    &     \\
Df4   & 0.952 (2.28e-5)    & 0.964 (7.32e-6)    & 0.976 (1.47e-6)    & 1.00 (0.00)        &     \\
Df5   & 0.976 (1.47e-6)    & 0.988 (9.31e-8)    & 0.988 (9.31e-8)    & 0.952 (2.28e-5)    & 1.00 (0.00) \\
\bottomrule
\end{tabular}
\end{table}

\begin{table}[htbp]
\caption{GPT’s Spearman’s Rank Correlation Matrix: Laptops}
\label{tab:gpt-laptops}
\centering
\footnotesize 
\renewcommand{\arraystretch}{1.3}
\begin{tabular}{lccccc}
\toprule
      & Df1               & Df2               & Df3               & Df4               & Df5               \\
\hline
Df1   & 1.00 (0.00)       &                   &                   &                   &                   \\
Df2   & 1.00 (0.00)       & 1.00 (0.00)       &                   &                   &                   \\
Df3   & 1.00 (0.00)       & 1.00 (0.00)       & 1.00 (0.00)       &                   &                   \\
Df4   & 1.00 (0.00)       & 1.00 (0.00)       & 1.00 (0.00)       & 1.00 (0.00)       &                   \\
Df5   & 0.988 (9.31e-8)   & 0.988 (9.31e-8)   & 0.988 (9.31e-8)   & 0.988 (9.31e-8)   & 1.00 (0.00)       \\
\bottomrule
\end{tabular}
\end{table}

\begin{table}[htbp]
\caption{DeepSeek’s Spearman’s Rank Correlation Matrix: Laptops}
\label{tab:deepseek-laptops}
\centering
\footnotesize 
\renewcommand{\arraystretch}{1.3}
\begin{tabular}{lccccc}
\toprule
      & Df1               & Df2               & Df3               & Df4               & Df5               \\
\hline
Df1   & 1.00 (0.00)       &                   &                   &                   &                   \\
Df2   & 0.988 (9.31e-8)   & 1.00 (0.00)       &                   &                   &                   \\
Df3   & 0.988 (9.31e-8)   & 1.00 (0.00)       & 1.00 (0.00)       &                   &                   \\
Df4   & 0.988 (9.31e-8)   & 0.976 (1.47e-6)   & 0.976 (1.47e-4)   & 1.00 (0.00)       &                   \\
Df5   & 0.976 (1.47e-6)   & 0.988 (9.31e-8)   & 0.988 (9.31e-8)   & 0.964 (7.32e-6)   & 1.00 (0.00)       \\
\bottomrule
\end{tabular}
\end{table}

\FloatBarrier

\end{document}